\documentclass[a4paper,12pt]{article}
\usepackage{amsfonts}
\usepackage{amsmath}
\usepackage{amssymb}
\usepackage{latexsym}
\usepackage{color}
\usepackage{colordvi}
\usepackage{epsfig}
\usepackage{array}
\usepackage{pifont}

\newcommand{\beq}{\begin{eqnarray}}
\newcommand{\eeq}{\end{eqnarray}}

\newcommand{\be}{\begin{equation}}
\newcommand{\ee}{\end{equation}}
\newcommand{\lwrsim}{\raise0.3ex\hbox{$<$\kern-0.75em\raise-1.1ex\hbox{$\sim$}}}
\newcommand{\lgrsim}{\raise0.3ex\hbox{$>$\kern-0.75em\raise-1.1ex\hbox{$\sim$}}}

\def\Am#1#2#3{\widetilde A_{#1}^{#2}(#3)}

\def\C2#1#2{({\cal C}_2)_{#1}^{#2}}

\def\Eq#1{Eq.~(\ref{#1})}

\def\eq#1{eq.~(\ref{#1})}

\def\prd#1#2#3{Phys.\ Rev.\ {\bf D#1} (#2) #3}
\def\npb#1#2#3{Nucl.\ Phys.\ {\bf B#1} (#2) #3}
\def\plb#1#2#3{Phys.\ Lett.\ {\bf B#1} (#2) #3}

\usepackage{geometry}
\usepackage{amsxtra}
\usepackage{axodraw}
\usepackage{cite}

\newcommand{\ghvertex}{\begin{picture}(100,25)(0,-3)
\SetWidth{1.2}
\DashArrowLine(12.5,0)(50,0){5}
\DashArrowLine(50,0)(87.5,0){5}
\Gluon(50,0)(50,25){-4}{3}
\CCirc(50,0){5}{Black}{Yellow}
\Text(12.5,-10)[l]{k}
\Text(87.5,-10)[r]{q}
\Text(60,20)[l]{q-k}
\end{picture}}
\newcommand{\ghost}{\begin{picture}(150,25)(0,0)
\SetWidth{1.2}
\DashArrowLine(12.5,0)(37.5,0){5}
\DashArrowLine(37.5,0)(112.5,0){5}
\DashArrowLine(112.5,0)(137.5,0){5}
\SetWidth{1}
\Vertex(37.5,0){2}
\Vertex(112.5,0){2}
\GlueArc(75,0)(37.5,0,90){-4}{6}
\GlueArc(75,0)(37.5,90,180){-4}{6}
\CCirc(75,37.5){10}{Black}{Blue}
\end{picture}}
\newcommand{\gluonTwoA}{\begin{picture}(150,25)(0,0)
\SetWidth{1.2}
\Gluon(12.5,0)(37.5,0){-4}{2}
\Gluon(37.5,0)(112.5,0){-4}{6}
\Gluon(112.5,0)(137.5,0){-4}{2}
\SetWidth{1}
\Vertex(37.5,0){2}
\Vertex(112.5,0){2}
\GlueArc(75,0)(37.5,0,90){-4}{6}
\GlueArc(75,0)(37.5,90,180){-4}{6}
\CCirc(75,37.5){10}{Black}{Blue}
\end{picture}}
\newcommand{\gluonTwoB}{\begin{picture}(150,25)(0,0)
\SetWidth{1.2}
\Gluon(12.5,0)(75,0){-4}{6}
\Gluon(75,0)(137.5,0){-4}{6}
\SetWidth{1}
\Vertex(75,0){3}
\GlueArc(75,20)(20,-90,90){4}{6}
\GlueArc(75,20)(20,90,270){4}{6}
\CCirc(75,40){10}{Black}{Blue}
\end{picture}}

\title{Ghost-gluon running coupling, power corrections and 
the determination of $\Lambda_{\overline {\rm MS}}$}
\author{Ph.~Boucaud$^a$, F. De soto$^b$, J.P.~Leroy$^a$, A.~Le~Yaouanc$^a$ \\
J. Micheli$^a$, O. P\`ene$^a$, J.~Rodr\'iguez-Quintero$^c$}
\date{}

\begin{document}
\maketitle

\begin{center}
$^a$Laboratoire de Physique Th\'eorique et Hautes
Energies\footnote{Unit\'e Mixte de Recherche 8627 du Centre National de
la Recherche Scientifique}\\
{Universit\'e de Paris XI, B\^atiment 211, 91405 Orsay Cedex,
France}\\
$^b$ Dpto. Sistemas F\'isicos, Qu\'imicos y Naturales, \\
Universidad Pablo de Olavide, 41013 Sevilla, Spain.
\\
$^c$ Dpto. F\'isica Aplicada, Fac. Ciencias Experimentales,\\
Universidad de Huelva, 21071 Huelva, Spain.
\end{center}


\begin{abstract}

We compute a formula including OPE power corrections to describe the
running of a QCD coupling non-perturbatively defined through the
ghost and gluon dressing functions. This turns out to be rather accurate.
We propose the ``{\it plateau}''-procedure to compute
$\Lambda_{\overline{\rm MS}}$ from the lattice computation of the running
coupling constant.  We show  a good agreement  between the different methods which
have been used to estimate $\Lambda_{\overline{\rm MS}}^{N_f=0}$.
We argue that $\Lambda_{\overline{\rm MS}}$ or the strong coupling constant
computed with different lattice spacings may be used to estimate the lattice spacing ratio.

\end{abstract}

\begin{flushleft}
LPT-Orsay 08-91\\
UHU-FT/08-10
\end{flushleft}

\section{Introduction}

Much work has been devoted in the last years to the study of the QCD running 
coupling constant determined from lattice simulations, 
as well in its perturbative 
regime~\cite{Bali:1992ru,Luscher:1993gh,deDivitiis:1994yp,Alles:1996ka,Boucaud:1998bq,Boucaud:2000ey,OPEtree,OPEone,Sternbeck:2007br} 
as in the deep infrared domain~\cite{Boucaud:2002nc}. 
The two main approaches to obtain the running coupling in terms of the renormalization 
momentum were either an application of the Schr\"odinger functional method with special boundary 
conditions or the confrontation of  the behaviour with respect to the renormalization scale 
of 2-gluon and 3-gluon Green functions with the corresponding perturbative predictions. The latter 
Green's functions approach also revealed a dimension-two non-zero gluon condensate in the landau gauge. 
Much work has been also done to investigate its 
phenomenological implications in the gauge-invariant world~\cite{Gubarev:2000nz}. 
In a very recent work~\cite{Sternbeck:2007br}, 
the Green's function approach to estimate $\Lambda_{\overline{\rm MS}}$ has been 
pursued by exploiting a non-perturbative definition of the coupling derived from the 
ghost-gluon vertex and computed over a large momentum window 
in the perturbative regime.
Much of this work was based on the analysis of {\it quenched} lattice simulations and led to the 
determination of $\Lambda_{\overline{\rm MS}}$ in pure Yang-Mills ($N_f=0$). 
Works on {\it unquenched} lattice configurations ($N_f=2$) started some time ago \cite{Boucaud:2001qz} 
and have been more actively pursued recently.  

Many unquenched configurations are presently available and
we are planing to apply what we have learned on pure Yang-Mills to gauge configurations with
twisted $N_f=2$~\cite{Boucaud:2008xu}) and $N_f=2+1+1$ dynamical quarks.
Thus, a very realistic estimate of $\Lambda_{\overline{\rm MS}}$, directly comparable with 
experimental determinations, will become an immediate possibility.
With the latter remarks in mind, we pay attention in this paper to study the above-mentioned 
non-perturbative coupling derived from the ghost-gluon vertex for being applied to the analysis of 
lattice data. We show in section 2 
that,  when the incoming ghost-momentum vanishes --and only in this case--  this ghost-gluon 
vertex can be directly related to the bare gluon and ghost propagators; we then obtain 
a formula to describe its running including non-perturbative power corrections. 
We propose to confront this formula  with lattice estimates of the coupling  and argue that this constitutes an 
optimal method for the identification of $\Lambda_{\overline{\rm MS}}$ and of  
the gluon condensate. In particular, it benefits of two main advantages: to have only two-points function 
to deal with (much simpler to be managed and more precise than three-points ones) and that the precision 
could be improved by extending the analysis of lattice data over a very large momenta window.
In section 3, we apply this procedure to previouly published lattice data 
for quenched simulations with a two-sided goal: (i) to check the method and (ii) to confirm 
the consistency of the picture we have acquired for the UV behaviour of Green functions 
in pure Yang-Mills. We finally conclude in section 4.

\section{The ghost-gluon coupling}

 There is a large number of possibilities to define the QCD renormalized coupling constant, depending on the observable used to measure it and on the renormalization scheme. Actually, any observable which behaves, from the perturbative point of view, as $g$ provides a suitable definition for it. Among such quantities stand the 3-gluon and the ghost-gluon vertices, which have been widely used by the lattice community to get a direct knowledge of $\alpha_s$ from  simulations. Of  course an important criterion to choose among those definitions will be how easy it is to connect it to other commonly used definitions, specially the $\overline{MS}$ one, and to extract from  it fundamental parameters like $\Lambda_{QCD}$.

A convenient class of renormalization schemes to work with on the lattice is made of the 
so-called ``$MOM$" schemes which are defined through the requirement that a given scalar coefficient function of the Green's function under consideration take  its tree-level value in a specific kinematical situation given  up to an overall ``renormalization scale" . To make the point clearer we recall 2 schemes which we have used in previous works on $\alpha_s$:
\begin{itemize}
\item The symmetric 3-gluon scheme in which one uses the 3-gluon vertex $\Gamma_{\mu\nu\rho}(p_1,p_2,p_3)$ with $p_1^2=p_2^2=p_3^2=\mu^2$
\item The asymmetric 3-gluon scheme ($\widetilde{MOM}$) in which  the 3-gluon vertex $\Gamma_{\mu\nu\rho}(p_1,p_2,p_3)$ is used with $p_1^2=p_2^2=\mu^2,\,p_3^2=0$
\end{itemize}

In the present note we shall apply a specific $MOM$-type renormalization scheme defined 
by fixing the (ghost and gluon) propagators and the ghost-gluon vertex at the renormalization point.
Let us start by writing the ghost  and gluon propagators  in Landau gauge as follows,
\beq
\left( G^{(2)} \right)_{\mu \nu}^{a b}(p^2,\Lambda) &=& \frac{G(p^2,\Lambda)}{p^2} \ \delta_{a b} 
\left( \delta_{\mu \nu}-\frac{p_\mu p_\nu}{p^2} \right) \ ,
\nonumber \\
\left(F^{(2)} \right)^{a,b}(p^2,\Lambda) &=& - \delta_{a b} \ \frac{F(p^2,\Lambda)}{p^2} \ ;
\eeq
$\Lambda$ being some regularisation parameter ($a^{-1}(\beta)$ if, for instance, we specialise to lattice
regularisation). The renormalized dressing functions, $G_R$ and $F_R$ are defined through :
\begin{flushleft}
\begin{align} \label{bar}
G_R(p^2,\mu^2)\ &= \ \lim_{\Lambda \to \infty} Z_3^{-1}(\mu^2,\Lambda) \ G(p^2,\Lambda)\nonumber\\
F_R(p^2,\mu^2)\ &= \ \lim_{\Lambda \to \infty} \widetilde{Z}_3^{-1}(\mu^2,\Lambda)\ F(p^2,\Lambda) \ ,
\end{align}
\end{flushleft}
\noindent with renormalization condition
\beq\label{bar2}
G_R(\mu^2,\mu^2)=F_R(\mu^2,\mu^2)=1 \ .
\eeq
Now, we will consider the ghost-gluon vertex which could be non-perturbatively obtained through 
a three-point Green function, defined by two ghost and one gluon fields, 
with amputated legs after dividing by two ghost and one gluon 
propagators. This vertex can be written quite generally as:

\beq\label{defGamma}
\widetilde{\Gamma}^{abc}_\nu(-q,k;q-k) = 
\ghvertex =
i g_0 f^{abc} 
\left( q_\nu H_1(q,k) + (q-k)_\nu H_2(q,k) \right) \ ,
\eeq

\noindent where $q$ is the outgoing ghost momentum and $k$ the incoming one, 
and renormalized according to:
\beq
\widetilde{\Gamma}_R=\widetilde{Z}_1 \Gamma.
\eeq
\noindent The vertex $\Gamma_\nu$  involves two independent scalar functions. In the MOM renormalization 
procedure $\widetilde{Z}_1$ is fully determined by demanding that one specific combination of those two form factors 
(chosen at one's will) be equal to its tree-level value for a specific kinematical 
configuration. 
We choose to apply MOM prescription for the scalar function $H_1+H_2$ that multiplies $q_\nu$ in \eq{defGamma} and 
the renormalization condition reads\footnote{In the case of zero-momentum gluon, an appropriate choice would be 
$\widetilde{Z}_1(\mu^2) H_1(q,q)|_{q^2=\mu^2}=1$. 
This would make the renormalized vertex equal to its tree-level value at 
the renormalization scale.} 
\beq\label{MOMT}
\left.(H^R_1(q,k) +  H^R_2(q,k))\right\vert_{q^2=\mu^2} = 
\lim_{\Lambda \to \infty}\widetilde{Z}_1(\mu^2,\Lambda)\left.(H_1(q,k;\Lambda) 
+  H_2(q,k;\Lambda))\right\vert_{q^2=\mu^2} =1,
\eeq
where we prescribe a kinematics for the substraction point such that 
the outgoing ghost momentum is evaluated at the renormalization scale, while the incoming one, $k$, depends on the 
choice of several possible configurations; for instance: $k^2=(q-k)^2=\mu^2$ (symmetric configuration) or 
$k=0, \ (q-k)^2=\mu^2$ (asymmetric-ghost configuration). 

On the other hand, the fields involved in the non-perturbative definition of the 
vertex $\Gamma_\nu$ in \eq{defGamma} can be directly renormalized by their renormalization constants, 
$Z_3$ and $\widetilde{Z}_3$, 
and the same MOM prescription applied to the scalar combination $H_1+H_2$ 
also implies:
\beq\label{g2R}
g_R(\mu^2) &=& \lim_{\Lambda \to \infty} \ \widetilde{Z}_3(\mu^2,\Lambda) Z_3^{1/2}(\mu^2,\Lambda) g_0(\Lambda^2)   
\left. \left(  H_1(q,k;\Lambda) + H_2(q,k;\Lambda) 
\rule[0cm]{0cm}{0.5cm}  \right) \right|_{q^2 \equiv \mu^2} 
\nonumber \\
&=&  \ \lim_{\Lambda \to \infty} g_0(\Lambda^2) \ 
\frac{Z_3^{1/2}(\mu^2,\Lambda^2)\widetilde{Z}_3(\mu^2,\Lambda^2)}{ \widetilde{Z}_1(\mu^2,\Lambda^2)} \ .
\eeq
We combine both \eq{MOMT} and the first-line equation of (\ref{g2R}) to replace $H_1+H_2$ and obtain 
the second line that shows the well-known relationship $Z_g=(Z_3^{1/2} \widetilde{Z}_3)^{-1} \widetilde{Z}_1$, 
where $g_R=Z_g^{-1} g_0$.

We turn now to the specific $MOM$-type renormalization scheme defined by a {\bf{zero incoming ghost momentum}}. 
Since those kinematics are the ones (and the only ones) in which Taylor's well known non-renormalization theorem 
(cf. ref~\cite{Taylor}) is valid we shall refer to this scheme as to the $T$-scheme and the corresponding 
quantities will bear a $T$ subscript. Then, in  eq~(\ref{defGamma}), we set $k$ to $0$ and get 
\beq\label{defGammaT}
\widetilde{\Gamma}^{abc}_\nu(-q,0;q) = 
i g_0 f^{abc} 
\left(H_1(q,0) +  H_2(q,0) \right)\, q_\nu \ .
\eeq
Now, Taylor's theorem states that $H_1(q,0;\Lambda) +  H_2(q,0;\Lambda)$ is equal to 1 in full QCD for 
any value of $q$. Therefore, the renormalization condition \eq{MOMT} implies $\widetilde{Z}_1(\mu^2)=1 $ and then 
\beq\label{alpha} 
\alpha_T(\mu^2) \equiv \frac{g^2_T(\mu^2)}{4 \pi}=  \ \lim_{\Lambda \to \infty} 
\frac{g_0^2(\Lambda^2)}{4 \pi} G(\mu^2,\Lambda^2) F^{2}(\mu^2,\Lambda^2) \ ;
\eeq
where we also apply the renormalization condition for the propagators, eqs. (\ref{bar},\ref{bar2}), 
to replace the renormalization constants, $Z_3$ and $\widetilde{Z}_3$, by the bare dressing 
functions. The remarkable feature of  \eq{alpha} is that  it involves only $F$ and $G$ so that 
no measure of the ghost-gluon vertex is needed for the determination of the coupling constant.

Equation~(\ref{alpha}) has extensively been advocated and studied on the lattice 
(see for instance reference~\cite{von Smekal:1997is}) and used for a determination 
of $\Lambda_{QCD}$ in reference~\cite{Sternbeck:2007br}. However it must be stressed 
that the $T$-scheme is the {\bf only} one in which  $\widetilde{Z_1}=1$. In any other 
scheme $\widetilde{Z_1}$ will be finite (since going from one scheme to any other one only 
involves an additional finite renormalization) but will keep a non trivial dependence 
on the scale, in particular for the symmetric scheme of reference~\cite{Seidensticker} that
has been computed at one loop in ref.~\cite{unpublished}. In such cases one must 
in principle apply the general definition (\ref{g2R}) of the coupling constant; 
nevertheless the form (\ref{alpha}) is used quite often in this case (for a 
kinematical configuration other than T-scheme's) also as an approximation, 
specially in relation with the study of Dyson-Schwinger equations.

We conclude this section by recalling that, in any scheme, the standard renormalization flow
dictating the evolution  with respect to the scale,
\beq\label{g2R-flow}
g^2_R(\mu^2) 
&=&
g^2_R(\mu^{\prime 2}) \ 
\left( \frac{\widetilde{Z}_1(\mu^{\prime 2})}{\widetilde{Z}_1(\mu^2)} \right)^2
F^2_R(\mu^2,\mu^{\prime 2}) G_R(\mu^2,\mu^{\prime 2}) \ ,
\eeq
will be straightfowrdly obtained from the second line of \eq{g2R} and the propagators 
renormalization conditions in eqs.~(\ref{bar},\ref{bar2}), where
\beq
\widetilde{Z}_1(\mu^2) = \lim_{\Lambda \to \infty} \widetilde{Z}_1(\mu^2,\Lambda^2)
\eeq
because of the Taylor's non-renormalization theorem. 
Of course, \eq{g2R-flow} reduces to 
\beq\label{g2RT}
g^2_T(\mu^2) 
&=&
g^2_T(\mu^{\prime 2}) 
F^2_R(\mu^2,\mu^{\prime 2}) G_R(\mu^2,\mu^{\prime 2}) \ .
\eeq
in the $T$-scheme.

\subsection{Pure perturbation theory}
\label{PTh}

In ref.~\cite{Chetyrkin00}, the three-loop perturbative substraction of all the 
three-vertices appearing in the QCD Lagrangian for kinematical configurations with one 
vanishing momentum has been done (in particular, the one involved in the definition of 
the coupling by \eq{alpha}). Different definitions of the coupling constant can be related  in perturbation theory through relations like :
\beq\label{alpha2}
\alpha_T(\mu^2) \ = \ \overline{\alpha}(\mu^2) \ \left( 1 + \sum_{i=1} c_i 
\left( \frac{\overline{\alpha}(\mu^2)}{4 \pi}\right)^i \ \right) \ ;
\eeq
on the other hand, since \eq{alpha} completely defines the running of the coupling,  after properly 
deriving  both its l.h.s. and r.h.s., one obtains
\beq\label{betah}
\frac{1}{\alpha_T(\mu^2)} \ \frac{d\alpha_T(\mu^2)}{d\overline{\alpha}} & = & 
\frac{1}{\beta_{\overline{\rm MS}}(\overline{\alpha})} 
\left( 2 \ \lim_{\Lambda \to \infty} \frac d {d\ln{\mu^2}} \ln F(\mu^2,\Lambda)
+ \lim_{\Lambda \to \infty} \frac d {d\ln{\mu^2}} \ln G(\mu^2,\Lambda)
\right)
\nonumber 
\\
&=&  \frac{2 \widetilde{\gamma}(\overline{\alpha}) +\gamma(\overline{\alpha})}
{\beta_{\overline{\rm MS}}(\overline{\alpha})}
\rule[0.5cm]{0cm}{0.5cm} \ \ ;
\eeq
where 
\beq\label{betaMS}
\beta_{\overline{\rm MS}}(\overline{\alpha}) \ = \ 
\frac{d\overline{\alpha}}{d\ln{\mu^2}} \ = \ - 4 \pi \ 
\sum_{i=0} \overline{\beta}_i \left( \frac{\overline{\alpha}} {4 \pi} \right)^{i+2} 
\eeq
is the standard $\beta$-function for the running coupling renormalized according to the usual 
$\overline{\rm MS}$ prescription,  while
\beq\label{gammas}
\widetilde{\gamma}(\overline{\alpha}) \ = \ 
\lim_{\Lambda \to \infty} \frac{d\ln{\widetilde{Z}_{3,{\rm MOM}}(\mu^2,\Lambda)}}{d\ln{\mu^2}} 
\ = \ 
\lim_{\Lambda \to \infty} \frac{d\ln{F(\mu^2,\Lambda)}}{d\ln{\mu^2}} 
\ = \ - \sum_{i=0} \widetilde{\gamma}_i \left( \frac{\overline{\alpha}}{4 \pi} \right)^{i+1} 
\nonumber \\
\gamma(\overline{\alpha}) \ = \ 
\lim_{\Lambda \to \infty} \frac{d\ln{Z_{3,{\rm MOM}}(\mu^2,\Lambda)}}{d\ln{\mu^2}} \ = \ 
\lim_{\Lambda \to \infty} \frac{d\ln{G(\mu^2,\Lambda)}}{d\ln{\mu^2}} 
\ = \ - \sum_{i=0} \gamma_i \left( \frac{\overline{\alpha}}{4 \pi} \right)^{i+1}
\eeq
are the anomalous dimensions for gluon and ghost propagators, both renormalized along
MOM prescriptions ({\it i.e.}, $G_R(\mu^2,\mu^2)=F_R(\mu^2,\mu^2)=1$), but expanded 
in terms of the $\overline{\rm MS}$ coupling $\overline{\alpha}$. The $\overline{\beta}_i$ coefficients  in \eq{betaMS} have been computed up to four loops in ref.~\cite{Larin}, 
$\overline{\beta}_0$ and $\overline{\beta}_1$ 
being scheme-independent. Then, eqs.(\ref{alpha2},\ref{betaMS},\ref{gammas}) 
can be applied to \eq{betah} and one is led to deal with a coupled system of $n$ algebraic equations  
to compute the coefficients $c_i$ and determine $\alpha_T$ at $n$ loops. To summarize, 
the running of coupling constant $\alpha_T$, although formally defined from a three-point 
Green function, can be derived from the knowledge of the standard $\overline{\rm MS}$ 
$\beta$-function and only two-points functions for ghost and gluon.
These two anomalous dimensions were computed in the $\overline{\rm MS}$ scheme at four 
loops in ref.~\cite{Chetyrkin:2004mf} and were converted into the MOM scheme in ref.~\cite{Boucaud:2005gg} 
for $N_f=0$ by applying
\beq
  \label{eq:evolution}
\gamma_{\Gamma,MOM}(\overline{\alpha})
 &=&
  \lim_{\Lambda\rightarrow\infty}
  \frac{d\ln\left(Z_{\Gamma,{\rm MS}}(\mu^2,\Lambda)\right)}
  {d\ln \mu^{2}} + \frac{d\ln\left(\Gamma_{R,\overline{\rm MS}}(\overline{\alpha})\right)}
  {d\ln \mu^{2}} 
\nonumber \\
   &\equiv& \gamma_{\Gamma,\overline{\rm MS}}(\overline{\alpha})
   + \frac{d \; \overline{\alpha}}{d \ln\mu^{2}} \ 
   \frac{\partial}{\partial \; \overline{\alpha}} \ln\Gamma_{R,\overline{\rm MS}}(\overline{\alpha}) \ ,
\eeq 
where $\Gamma$ stands generically for the two bare two-point dressing functions $F$ and $G$, $\Gamma_R$ for the renormalized 
ones~\footnote{The gluon and ghost renormalized propagators in the $\overline{\rm MS}$ scheme were also provided by 
ref.~\cite{Chetyrkin00}} and $Z_\Gamma$ for the appropriate renormalization constant. 
\Eq{eq:evolution} provides also the coefficients $\widetilde{\gamma}_i$ and $\gamma_i$ 
for any $N_f$ (see appendix \ref{appendix}). 
Thus, one can solve the above mentioned coupled system of algebraic equations 
and obtain the coefficients $c_i$ in \eq{alpha2}, the first of those equations 
(the one stemming from matching the $1/\alpha$-terms  in the 2 sides) resulting 
in the following constraint~\footnote{\Eq{wellknown} is a well-known relation verified by scheme-independent 
coefficients of the ghost and gluon anomalous dimensions and of the  $\beta$-function.}:
\beq\label{wellknown}
2 \widetilde{\gamma}_0 + \gamma_0 = \overline{\beta}_0 \ ,
\eeq
which, in this context, results from \eq{alpha}. The three first coefficients $c_i$ in Landau gauge, 
for instance, will be:
\beq
c_1 &=& \frac{507-40 N_f}{36} \ ,
\nonumber \\
c_2 &=& \frac{76063}{144} - \frac{351} 8 \zeta(3) - 
    \left( \frac{1913}{27} + \frac 4 3 \zeta(3) \right) \ N_f 
    + \frac{100}{91} N_f^2 
\nonumber \\
c_3 &=& \frac{42074947}{1728} - \frac{60675}{16}\zeta(3) 
- \frac{70245}{64} \zeta(5) - \left( \frac{769387}{162} - 
\frac{8362}{27} \zeta(3) -\frac{2320}{9} \zeta(5) \right) \ N_f 
\nonumber \\
&+& \left( \frac{199903}{972} + \frac{28}{9} \zeta(3) \right) \ N_f^2
- \frac{1000}{729} \ N_f^3
\ .
\eeq
These three coefficients obviously define unambigously the running of $\alpha_T$ given in \eq{alpha} up to 
four-loops. In other words, one obtains for the $\beta$-function of $\alpha_T$,
\beq\label{beta}
\beta_T(\alpha_T) \ = \ 
\frac{d\alpha_T}{d\ln{\mu^2}} \ = \ - 4 \pi \ 
\sum_{i=0} \widetilde{\beta}_i \left( \frac{\alpha_T} {4 \pi} \right)^{i+2} \ ,
\eeq
the following coefficients up to four-loops
\beq\label{betacoefs}
\widetilde{\beta}_0 &=& \overline{\beta}_0 = 11 - \frac 2 3 N_f 
\nonumber \\
\widetilde{\beta}_1 &=& \overline{\beta}_1 = 102 - \frac{38} 3 N_f 
\nonumber \\
\widetilde{\beta}_2 &=& \overline{\beta}_2 -\overline{\beta}_1 c_1 + \overline{\beta}_0 (c_2-c_1^2) 
\nonumber \\
&=&3040.48 \ - \ 625.387 \ N_f \ + \ 19.3833 \ N_f^2 
\nonumber \\
\widetilde{\beta}_3 &=& \overline{\beta}_3 - 2 \overline{\beta}_2 c_1 + \overline{\beta}_1 c_1^2 
+ \overline{\beta}_0 (2 \ c_3 - 6 \ c_2 c_1 + 4 \ c_1^3) 
\nonumber \\
&=&  100541 \ - \ 24423.3 \ N_f \ + \ 1625.4 \ N_f^2 \ - \ 27.493 \ N_f^3 
\ ,
\eeq
 These coefficients $\widetilde{\beta}_i$ are the same as the ones obtained in 
ref.~\cite{Chetyrkin00} thanks to a direct application of the MOM prescription to the ghost-gluon 
coupling with vanishing incoming-ghost momentum, as it should be.
As for  the $\Lambda_{\rm QCD}$ parameters in the two schemes, they are 
related through
\beq\label{ratTMS}
\frac{\Lambda_{\overline{\rm MS}}}{\Lambda_T} \ = \ e^{\displaystyle -\frac{c_1}{2 \beta_0}} \ = \ 
e^{\displaystyle - \frac{507-40 N_f}{792 - 48 N_f}}
\ .
\eeq
\Eq{beta} can be integrated and perturbatively inverted to obtain the following standard four-loop formula for 
the running coupling:
\begin{align}
  \label{betainvert}
  \begin{split}
      \alpha_T(\mu^2) &= \frac{4 \pi}{\beta_{0}t}
      \left(1 - \frac{\beta_{1}}{\beta_{0}^{2}}\frac{\log(t)}{t}
     + \frac{\beta_{1}^{2}}{\beta_{0}^{4}}
       \frac{1}{t^{2}}\left(\left(\log(t)-\frac{1}{2}\right)^{2}
     + \frac{\widetilde{\beta}_{2}\beta_{0}}{\beta_{1}^{2}}-\frac{5}{4}\right)\right) \\
     &+ \frac{1}{(\beta_{0}t)^{4}}
 \left(\frac{\widetilde{\beta}_{3}}{2\beta_{0}}+
   \frac{1}{2}\left(\frac{\beta_{1}}{\beta_{0}}\right)^{3}
   \left(-2\log^{3}(t)+5\log^{2}(t)+
\left(4-6\frac{\widetilde{\beta}_{2}\beta_{0}}{\beta_{1}^{2}}\right)\log(t)-1\right)\right)
\\
%
%
&\mbox{\rm{with}} \ \ t=\ln{\frac{\mu^2}{\Lambda_T^2}} \ . 
     \end{split}
\end{align}
%

As a last remark, applying the approximation $\widetilde{Z}_1=1$ for symmetric 
(ghost-gluon vertex renormalized at a symmetric momenta configuration) or 
soft-gluon (vertex renormalized at a vanishing-gluon momenta configuration) schemes 
implies that the same lattice data for the coupling, obtained through \eq{alpha}, 
would be confronted to different perturbative formulae analogous to \eq{betainvert} with 
$\beta$-function coefficients and $\Lambda_{\rm QCD}$ parameters apropriate for each scheme. 
Thus, the systematic deviation induced by applying this approximation to 
the determination of $\Lambda_{\overline{\rm MS}}$ from the confrontation of 
perturbation theory and lattice data, provided that $\beta_0$ and $\beta_1$ 
are scheme-independent, mainly results from the ratio of $\Lambda_{\rm QCD}$ to 
$\Lambda_{\overline{\rm MS}}$ in \eq{ratTMS}. For instance in pure Yang-Mills, 
if one takes $N_f=0$ in \eq{ratTMS}, it gives a ratio of 0.527 in T-scheme, 
while the same ratio for instance in symmetric and soft-gluon schemes
is 0.463 (14 \% of error) and 0.429 (23 \% of error), respectively.

\subsection{OPE power corrections}
\label{OPEsection}

One of the goals of the present paper consists in obtaining a formula for the QCD running coupling
that could be implemented in conjunction with lattice estimates 
to determine a ``{\it plateau}'' for $\Lambda_{\rm QCD}$ in terms of 
the momentum, as will be explained in the next section. In order to extend this ``{\it plateau}''
to energies as low as possible (of the order of 3 GeV) and to take full advantage of  the lattice 
data in order to  reduce the systematic uncertainties, it is mandatory to take into account  the 
gauge-dependent dimension-two OPE power corrections (cf.~\cite{OPEtree,OPEone,Boucaud:2002nc,Dudal:2002pq}) 
to  $\alpha_T$.

The leading power contribution to the ghost propagator, 
\beq\label{GhProp}
(F^{(2)})^{a b}(q^2) = \int d^4x e^{i q \cdot x} 
\langle \ T\left( c^a(x) \overline{c^b}(0) \right) \ \rangle 
\eeq
can be computed using the operator product 
expansion~\cite{Wilson69} (OPE), as is done in ref.~\cite{Boucaud:2005xn},
\beq\label{GhExp}
T\left( c^a(x)  \overline{c^b}(0) \right) = \sum_t \left(c_t\right)^{a b}(x) \ O_t(0);
\eeq
here $O_t$ is a local operator, regular when $x \to 0$, and  the Wilson coefficient $c_t$ 
contains the short-distance singularity. \Eq{GhExp} involves a full hierarchy of terms, ordered according to their mass-dimension, among which only  ${\bf 1}$ and $:A_\mu^a A_\nu^b:$ contribute to \eq{GhProp} in Landau gauge \footnote{The operators with 
an odd number of fields ($d=1,3/2$; $\partial_\mu A$ and $\partial_\mu \overline{c}$) cannot satisfy colour 
and Lorentz invariance and do not contribute  a non-zero non-perturbative expectation value, 
and  $\overline{c} c$ does not contribute either  because of the particular tensorial structure of the ghost-gluon 
vertex.} up to the order $1/q^4$.
Then, using \eq{GhExp} into \eq{GhProp}, we obtain:
\beq\label{OPE1}
(F^{(2)})^{a b}(q^2) &=& (c_0)^{a b}(q^2) \ + \ \left( c_2 \right)^{a b \sigma \tau}_{s t}(q^2)
\langle : A_\sigma^s(0) A_\tau^t(0): \rangle \ + \ \dots \nonumber \\ 
&=& (F_{\rm pert}^{(2)})^{a b}(q^2) \ + \ 
w^{a b} \ \frac{\langle A^2 \rangle}{4 (N_C^2-1)} \ + \ \dots 
\eeq
where 
\beq\label{OPE3}
w^{a b} \ &=& \ \left( c_2 \right)^{a b \sigma \tau}_{s t} \delta^{s t} g_{\sigma \tau} \ = \ 
\frac 1 2 \ \delta^{s t} g_{\sigma \tau} \frac{\int d^4x e^{i q \cdot x} \
\langle \Am{\tau'}{t'}{0} \ T\left( c^a \overline{c^b} \right) \ \Am{\sigma'}{s'}{0} \rangle_{\rm connected}}
{{G^{(2)}}_{\sigma \sigma'}^{s s'} {G^{(2)}}_{\tau \tau'}^{t t'} } \nonumber \\
&=&   2 \times \rule[0cm]{0cm}{1.7cm} \ghost,
\eeq
and the SVZ factorisation~\cite{SVZ} is invoked to compute the Wilson coefficients. 
Thus, one should compute the ``{\it sunset}'' diagram of the last line of \eq{OPE3}, that binds
the ghost propagator to the gluon condensate (where the blue bubble means contracting the 
color and lorentz indices of the incoming legs with $1/2 \delta_{st} \delta_{\sigma \tau}$) 
to obtain the leading non-perturbative contribution 
(of course, the first Wilson coefficient gives trivially the perturbative propagator).

Finally,
\vspace*{0.5cm}
\beq\label{Fin1}
(F^{(2)}_R)^{a b}(q^2,\mu^2) \ = \ (F^{(2)}_{R,{\rm pert}})^{a b}(q^2,\mu^2) \
\left( 1 + \frac{3}{q^2} \ 
\frac{g^2_R \langle A^2 \rangle_{R,\mu^2}} {4 (N_C^2-1)} \right) \ + \ {\cal O}\left(g^4,q^{-4} \right)
\eeq
where the $A^2$-condensate is renormalised at the subtraction point $q^2=\mu^2$, 
according to the MOM scheme definition, by imposing 
the tree-level value to the Wilson coefficient at the renormalization point. 
As far as we do not need to deal with the anomalous dimension of the $A^2$ operator, 
we can factorise the tree-level ghost propagator. The ghost dressing function is then written as:
\beq\label{Z3fantome}
F_R(q^2,\mu^2) \ = \ F_{R, {\rm pert}}(q^2,\mu^2) \
\left(  1 + \frac{3}{q^2} \frac{g^2_R \langle A^2 \rangle_{R,\mu^2}} {4 (N_C^2-1)} \right) \ ,
\eeq
where the multiplicative correction to the purely perturbative $F_{R,{\rm pert}}$ is determined 
up to corrections of the order $1/q^4$ or $\ln{q/\mu}$ 
(the Wilson coefficient at the leading logarithm is computed in 
appendix \ref{appendix2}). 

We can handle in the same way (see refs.\cite{OPEtree,OPEone})  the OPE power correction to the gluon propagator and  
obtain 
\beq\label{OPEgl3}
w_{\mu\nu}^{a b} \ &=&  \gluonTwoB
+ \
 2 \times \rule[0cm]{0cm}{1.7cm} \gluonTwoA \
\nonumber \\
&=&\rule[0.5cm]{0cm}{0.5cm} \frac{3 g^2}{q^2} \ (G_{\rm pert}^{(2)})_{\mu\nu}^{ab}.
\eeq
Then, after renormalization, one gets
\beq\label{OPEgl1}
(G^{(2)}_R)_{\mu\nu}^{a b}(q^2,\mu^2)&=& (G_{R,\rm pert}^{(2)})_{\mu\nu}^{a b}(q^2,\mu^2) \ + \ 
\left(w_{\mu\nu}^{a b}\right)_{R,\mu^2} \ \frac{\langle A^2 \rangle_{R,\mu^2}}{4 (N_C^2-1)} \ + \ \dots 
\nonumber \\
&=&\ (G^{(2)}_{R,{\rm pert}})_{\mu\nu}^{a b}(q^2,\mu^2) \
\left( 1 + \frac{3}{q^2} \ 
\frac{g^2_R \langle A^2 \rangle_{R,\mu^2}} {4 (N_C^2-1)} \right) \ + \ {\cal O}\left(g^4,q^{-4} \right) 
\ \ .
\eeq
and an appropriate projection gives for the gluon dressing function :
\beq\label{Z3glue}
G_R(q^2,\mu^2) \ = \ G_{R, {\rm pert}}(q^2,\mu^2) \
\left(  1 + \frac{3}{q^2} \frac{g^2_R \langle A^2 \rangle_{R,\mu^2}} {4 (N_C^2-1)} \right) \ .
\eeq

Finally, putting together the defining relation \eq{alpha} and the results eqs.~(\ref{Z3fantome},\ref{Z3glue}) we get
\beq\label{alphahNP}
\alpha_T(\mu^2) &=& \lim_{\Lambda \to \infty} 
\frac{g_0^2}{4 \pi} F^2(\mu^2,\Lambda) G(\mu^2,\Lambda) \nonumber \\
&=& 
\overbrace{\lim_{\Lambda \to \infty} 
\frac{g_0^2}{4 \pi} F^2(q_0^2,\Lambda) G(q_0^2,\Lambda)}^{\displaystyle
\alpha^{\rm pert}_T(q_0^2)} \ 
F^2_R(\mu^2,q_0^2) \ G_R(\mu^2,q_0^2)
\nonumber \\
&=&
\underbrace{
\alpha^{\rm pert}_T(q_0^2) F^2_{R,{\rm pert}}(\mu^2,q_0^2) \ G_{R,{\rm pert}}(\mu^2,q_0^2) 
}_{\displaystyle \alpha^{\rm pert}_T(\mu^2)}
\ 
\left( 
 1 + \frac{9}{\mu^2} \frac{g^2_T(q_0^2) \langle A^2 \rangle_{R,q_0^2}} {4 (N_C^2-1)}
\right) \ ,
\eeq
where $q_0^2 \gg \Lambda_{\rm QCD}$ is some perturbative scale and
the $\beta$-function, and its coefficients in \eq{betacoefs}, of course describe the running 
of the perturbative part of the evolution, $\alpha_T^{\rm pert}$. 

The Wilson coefficient at the leading logarithm for the T-scheme MOM running coupling 
is presented in appendix \ref{appendix2}, where we also 
show that the inclusion of the logarithmic correction would induce no significant 
effect, provided that the coupling multiplying $A^2$ inside the bracket is taken to be 
renormalized also in T-scheme. 
Thus, for the sake of simplicity, \eq{alphahNP} will be applied for our analysis in 
the next section.

\section{Data Analysis}
\label{Dat-An}

In the following, we will first propose a ``{\it plateau}''-procedure 
exploiting \eq{alphahNP} to get a reliable estimate of the 
$\Lambda_{\rm QCD}$-parameter from the lattice  and we will apply it to previously 
published quenched lattice data~\cite{Boucaud:2005xn,Boucaud:2005gg} as a 
check of the method.

\subsection{The  ``{\it plateau}'' method}

The goal being to get a trustworthy  estimate of the 
$\Lambda_{\overline{\rm MS}}$-parameter, one could attempt to do it by inverting 
the perturbative formula \eq{betainvert} and  using in the {\it inverted} formula
 the lattice estimates of the running coupling obtained by means of  \eq{alpha} for as 
many lattice momenta as possible. Then, one should look for 
a ``{\it plateau}'' of $\Lambda_{\overline{\rm MS}}$ in terms of 
momenta in the high-energy perturbative regime (this was done 
with the coupling defined by the three-gluon vertex in \cite{Alles:1996ka,Boucaud:1998bq}). 
In the next subsection, fig.~\ref{plot-plateau}.(a) shows the estimates of 
$\Lambda_{\overline{\rm MS}}$ so calculated for the lattice data presented 
in ref.~\cite{Boucaud:2005xn,Boucaud:2005gg} over $9 \ \lwrsim p^2 \ \lwrsim \ 33$~GeV$^2$.

However, in order to take advantage of the largest possible momenta window one can use instead \eq{alphahNP}. In this way we shall hopefully be able to extend towards {\it low} momenta the region over which to look for the best possible values of
the gluon condensate and of $\Lambda_{\overline{\rm MS}}$~\footnote{This increases the statistics and reduces errors. It also avoids 
some possible systematic deviation appearing when lattice momentum components, in lattice units, approach 
 $\pi/2$ (Brillouin's region border).}.
In other words, one requires the best-fit to a constant of 
\beq
(x_i,y_i) &\equiv& \left( p^2_i,\Lambda(\alpha_i) \right) \ , \nonumber \\
{\rm with:} \ \ \ \ \ \alpha_i &=& \frac{\alpha_{\rm Latt}(p^2_i)}{\displaystyle 1+\frac{c}{p^2_i}} \ ;
\eeq
where $\Lambda(\alpha)$ is obtained by inverting the 
perturbative four-loop formula, \eq{betainvert}, 
and $c$ results from the best-fit (it appeared written in terms of the gluon condensate in 
\eq{alphahNP}~). Of course, $\Lambda(\alpha)$ reaches a ``{\it plateau}'' (if it does) behaving 
in terms of the momentum as a constant that we will take as our estimate of 
$\Lambda_{\overline{\rm MS}}$.

\subsection{Applying the method}

The lattice data that we will exploit here to check the 
method we have explained above  were previously presented in 
ref.~\cite{Boucaud:2005gg}. We refer to this work for all the details concerning the lattice 
implementation: algorithms, action, Faddeev-Popov operator inversion, etc.

The parameters of the whole set of simulations are described in 
table~\ref{setup} 

\begin{table}[ht]
\centering
\begin{tabular}{c|c||c|c}
\hline
$\beta$ & Volume & $a^{-1}$ (GeV) & Number of confs.
\\ \hline
$6.0$ &  $16^4$ & $1.96$ & $1000$
\\ \hline
$6.0$ &  $24^4$ & $1.96$ & $500$
\\ \hline
$6.2$ &  $24^4$ & $2.75$ & $500$
\\ \hline
$6.4$ &  $32^4$ & $3.66$ & $250$
\\ 
\hline
\end{tabular}
\caption{Run parameters of the exploited data~\cite{Boucaud:2005gg}.}
\label{setup}
\end{table}

\subsubsection{The scaling from different lattices}
\label{scaling}

It should first be noted that the scaling of \eq{alpha} from the several lattices 
we use  is indeed satisfactory. The prescription of 
taking the infinite cut-off limit in \eq{alpha} means in practice to have the 
lattice artifacts under control. This is in fact the case for UV ones. 
In particular, the hypercubic artifacts behaving as ${\cal O}(a^k \sum p_i^k)$ for the 
lattice propagators we analyze were cured, as explained in \cite{Boucaud:2005gg}, by exploiting 
the $H_4$-symmetry. 

As an indirect way of testing that scaling, we consider all the 
lattice propagators as functions of the momentum {\it measured in lattice units}, ({\it i.e.} with dimensionless momenta 
$p_{\rm Lat} = a(\beta) p$, where $a(\beta)$ is the lattice spacing in physical 
units at the particular bare lattice coupling $g_0^2= 6/\beta$), and determine the 
ratios of $a(\beta)$'s for the scaling to work. Then, still working in lattice units, the 
best-fit parameters to be obtained by applying the  ``{\it plateau}''-method 
will be $a(\beta)\, \Lambda_{\overline{\rm MS}} $ and 
$a^2(\beta) g^2_T \langle A^2 \rangle_R$, and the ratio of those best-fit parameters 
for different lattices will provide the ratio of the corresponding lattice spacings. 

\begin{table}[ht]
\centering
\begin{tabular}{|c|c||c|c|c|}
\hline
$\beta$ & Volume & $a(\beta)/a(6.2)$ (this work) & $a(\beta)/a(6.2)$ \cite{Bali:1992ru} & deviations (\%)
\\ \hline
$6.0$ &  $16^4$ & $1.368$ & $1.378$ & 0.7
\\ \hline
$6.0$ &  $24^4$ & $1.322$ & $1.378$ & 4.1
\\ \hline 
$6.2$ &  $24^4$ & $1$ & $1$ & 0
\\ \hline
$6.4$ &  $32^4$ & $0.768$ & $0.751$ & 2.2
\\ 
\hline
\end{tabular}
\caption{\small Comparison of lattice spacings ratios obtained by means of  the scaling of 
\eq{alpha} as explained in the text and of the string-tension method.}
\label{cali}
\end{table}

\begin{figure}[htb]
\begin{center}
\begin{tabular}{cc}
\includegraphics[width=8cm]{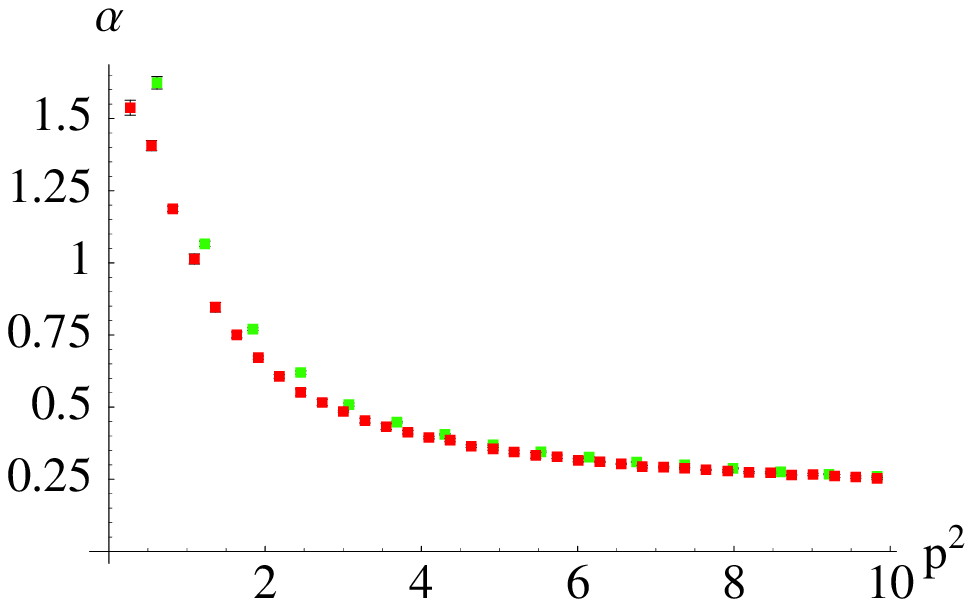} &
\includegraphics[width=8cm]{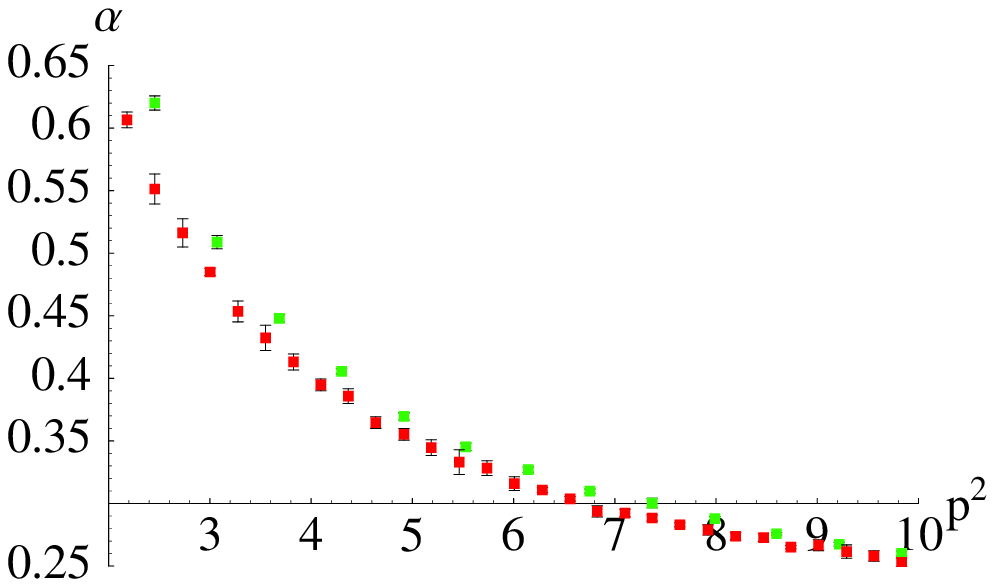}
\\ 
(a) & (b)
\end{tabular}
\end{center}
\caption{\small (a) Plot of $\alpha_T$ defined by \eq{alpha} in terms of the square of the 
renormalization momentum as computed from the two lattices at $\beta=6.0$ with different volumes:
$V=16^4$ (green boxes) and $V=24^4$ (red boxes). (b) A zoom onto the high momenta region of the left plot. } 
\label{volfini}
\end{figure}

In tab.~\ref{cali}, the ratio of lattice spacings obtained by the standard string-tension 
method~\cite{Bali:1992ru} are compared with those obtained as explained above. More precisely : (i) we first determine 
$\Lambda_{\overline{\rm MS}} a(6.2)$ and $a^2(6.2) g^2_T \langle A^2 \rangle_R$ for 
the lattice data with $\beta=6.2$; (ii) then, for each new $\beta$, we determine 
$x=a(\beta)/a(6.2)$ in such a way that a ``{\it plateau}'' for 
$x\, a(6.2) \,\Lambda_{\overline{\rm MS}}$  is obtained with a gluon condensate given 
by $ x^2\,a^2(6.2)\, g^2_T \langle A^2 \rangle_R$. They agree very well, at least for 
the ratios computed for the three lattice simulations with roughly the same physical 
volume: $\beta=6.0 (L=16=1.58$ fm), $\beta=6.2 (L=24=1.72$ fm), 
$\beta=6.4 (L=32=1.72$ fm). A slightly larger discrepancy ($\sim 4 \%$) appears when comparing 
with data for the largest lattice ($\beta=6.0, L=24=2.37$ fm).  We suspect  that this is the manifestation of 
a finite-volume effect.
Actually, if we compare the two simulations at $\beta=6.0$ for different volumes 
(see fig.~\ref{volfini}), such an effect can be seen, although 
it decreases as the physical momentum increases (and becomes in practice 
negligible at $p^2 \sim 9$ GeV$^2$). 

Thus, one can conclude that the scaling of the coupling defined by \eq{alpha} for $p^2 \lgrsim 9$ GeV$^2$ is very good. 
Conversely, this argument provides an alternative method 
to determine the lattice size for a simulation at a given $\beta$ in terms of the 
one known in physical units at any other one.

\subsubsection{Looking for the ``{\it plateau}''}

\begin{figure}[ht]
\begin{center}
\begin{tabular}{cc}
\begin{tabular}{c}
\includegraphics[width=7cm]{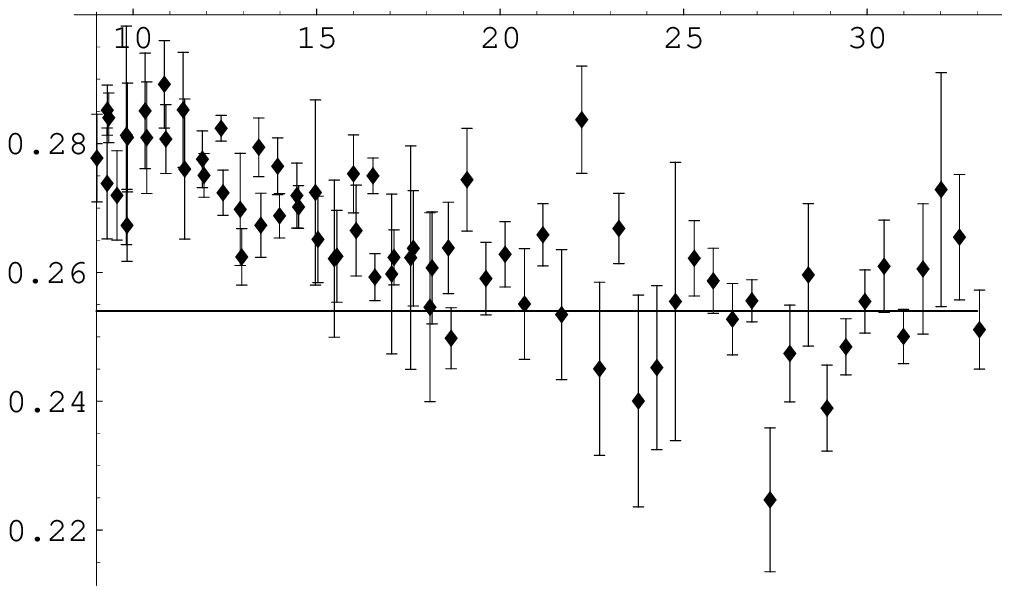}
\\  \rule[-1.9cm]{0cm}{3.8cm} (a) 
\end{tabular}
&
\begin{tabular}{c}
\includegraphics[width=7cm]{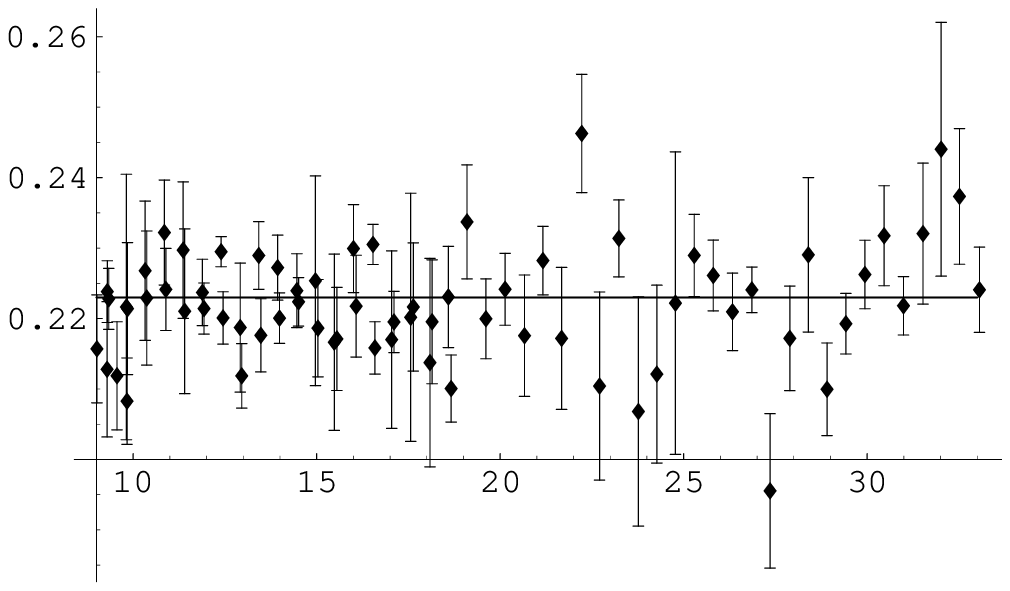}
\\ (b)
\end{tabular}
\end{tabular}
\vspace*{-2.1cm}
\end{center}
\caption{\small (a) Plot of  $\Lambda_{\overline{\rm MS}}$ (in GeV) computed by the inversion of 
the four-loop perturbative formula \eq{betainvert} as a function  of the square of the momentum (in GeV$^2$); 
 the coupling is estimated from the lattice data through \eq{alpha}.
(b) Same as plot (a) except for applying the non-perturbative formula \eq{alphahNP} for 
the coupling and looking for the gluon condensate generating the best plateau over 
$9 \ \lwrsim \ p^2 \ \lwrsim \ 33$ GeV$^2$.} 
\label{plot-plateau}
\end{figure}

In fig.~\ref{plot-plateau}.(a), we show  the estimates of $\Lambda_{\overline{\rm MS}}$ obtained when interpreting the lattice coupling computed by \eq{alpha} for any momentum 
$9 \ \lwrsim p^2 \ \lwrsim 33$ GeV$^2$ in terms of the 
 {\it inverted}  four-loop perturbative formula for the coupling, \eq{betainvert}. The estimates systematically decrease as 
the squared momentum increases until around 22 GeV$^2$;  above this value, only a 
noisy pattern results. In fig.~\ref{plot-plateau}.(b), the same is plotted but inverting  instead
the non-perturbative formula including power corrections, \eq{alphahNP}. The value of the gluon condensate has been determined 
by requiring a ``{\it plateau}'' to exist (as explained in the previous section) 
over the total momenta window.

\begin{figure}[htb]
\begin{center}
\begin{tabular}{cc}
\includegraphics[width=8cm]{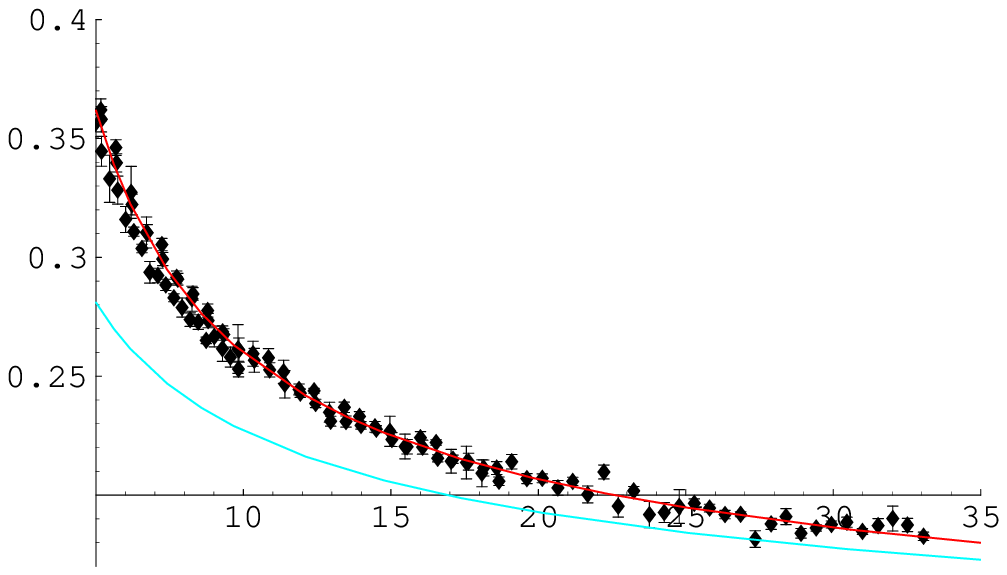} &
\includegraphics[width=8cm]{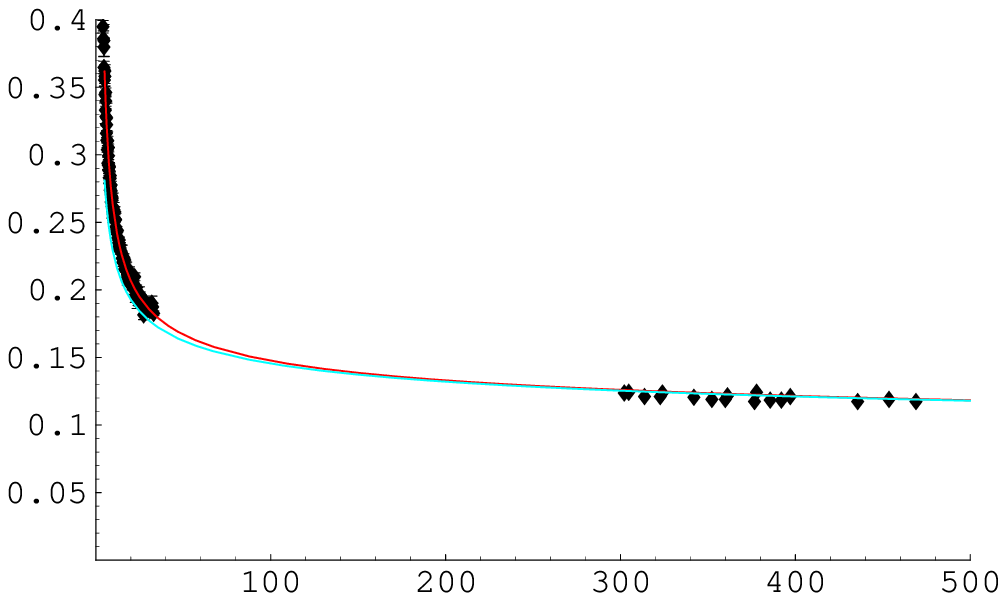}
\\ 
(a) & (b)
\end{tabular}
\end{center}
\caption{\small (a) Plot of $\alpha_T$ defined by \eq{alpha} in terms of the square of the 
renormalization momentum: the red solid line is computed with \eq{alphahNP}
with  $\Lambda_{\overline{\rm MS}}=224$ MeV, the blue one with \eq{betainvert} for the 
same $\Lambda_{\overline{\rm MS}}$ and the data are obtained from the lattice data 
set-up in table \ref{setup}. (b) The same but with some additional lattice estimates for 
the coupling at very high momenta (300--500 GeV$^2$) taken 
from \cite{Sternbeck:2007br}.}
\label{plot-alpha}
\end{figure}

One should realize that, had we not taken into account the noisy ballpark of points above 22 
GeV$^2$ and had we considered the perturbative regime as reached at that momentum, we would 
have got an estimate of $\Lambda_{\overline{\rm MS}}$ roughly 35-40 MeV above the one 
obtained from the non-perturbative formula. In other words, the non-perturbative 
analysis seems to indicate that the perturbative regime is far from being  achieved 
at $p=5$ GeV.  This is illustrated in figure~\ref{plot-alpha} in which, adopting for $\Lambda_{\overline{\rm MS}}$ the value $224$ MeV which results from the non-perturbative analysis, we plot against the square of the renormalization momentum the coupling constant as computed by means of the non-perturbative formula (\ref{alphahNP}) (red curve) and of the perturbative one  (\ref{betainvert}) (blue curve). Displayed are also the lattice data, {\it i.e.} the values of $\alpha_T$ obtained from \eq{alpha}. In 
figure~\ref{plot-alpha}.a the range in $\mu^2$ one sees that the non-perturbative approach provides a fairly good agreement with the data, the 
$\chi^2$ being 1.3 per degree of freedom. On the contrary there is a clear disagreement with the perturbative formula. Furthermore, one can extrapolate the 
value of the $\alpha_T$ up to very high momenta with \eq{alphahNP}, 
$p^2 \sim 300-500~\rm{GeV}^2$, where the purely perturbative \eq{betainvert} and 
the non-perturbative \eq{alphahNP}, both with the same $\Lambda_{\overline{\rm MS}}$, 
generate in practice the same results. The plot of fig.~\ref{plot-alpha}.(b) shows indeed that 
the curve for the coupling extrapolated in this way  joins perfectly the lattice estimates at 
high momenta taken from \cite{Sternbeck:2007br}. Thus, the inclusion of the non-perturbative 
OPE power correction, \eq{alphahNP}, to describe the running of the coupling 
eliminates effectively  the observed systematic deviations for the estimates 
of $\Lambda_{\overline{\rm MS}}$ from the momenta window from 3 GeV to 5 GeV 
(fig.~\ref{plot-plateau}.(a)~) and essentially leads to the same estimate as was found from 
the perturbative regime at very high momentum.


\begin{figure}[htb]
\begin{center}
\begin{tabular}{cc}
\includegraphics[width=8cm]{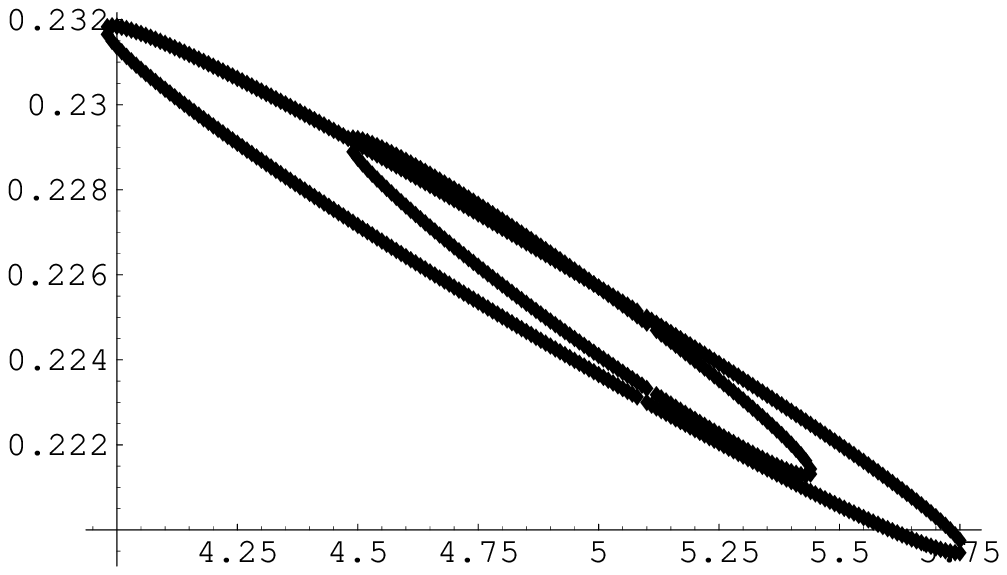}
&
\includegraphics[width=7cm]{lambdas-wav.eps}
\\ 
(a) & (b)
\end{tabular}
\end{center}
\caption{\small (a) The ellipsoid defined by 
$\chi^2(\Lambda_{\overline{\rm MS}},g^2_T \langle A^2 \rangle_R) = \chi^2_{\rm min} + 1$. 
The y-axis is for $\Lambda_{\overline{\rm MS}}$ expressed in GeV and 
x-axis for $g^2_T \langle A^2 \rangle_R$ in GeV$^2$. The small ellipsoid is obtained for 
a fitting window defined by $p^2 > 9$ GeV$^2$ and the larger is for 
$p^2 > 14$ GeV$^2$. (b) Comparison with previous estimates of $\Lambda_{\overline{\rm MS}}$
in pure Yang-Mills collected in tab.~\ref{comp}; the blue triangle stands for the estimate 
in this work and the red square for the {\it average} of the others. The 1-$\sigma$ error 
interval for the average (dashed red line) were estimated by treating the errors in 
tab.~\ref{comp} as purely statistical ones.}
\label{chi2eps}
\end{figure}

Thus, we have been able to obtain simultaneous best-fit values for both the gluon condensate 
and $\Lambda_{\overline{\rm MS}}$. It is however manifest that they are 
correlated by their determination: the larger  the gluon condensate is, the smaller 
 the value of $\Lambda_{\overline{\rm MS}}$ has to be. In fig.~\ref{chi2eps}, we plot 
the ellipsoid defined by~\footnote{The errors on the lattice estimates 
of the coupling that were used to compute $\chi^2$ were obtained by propagating 
the ones computed through  the  jackknife method for F and G in \cite{Boucaud:2005gg}.}  
$\chi^2(\Lambda_{\overline{\rm MS}},g^2_T \langle A^2 \rangle_R) = \chi^2_{\rm min} + 1$ 
for a fitting window defined by $p^2 > 8$ GeV$^2$ and for one restricted to $p^2 > 14$ 
GeV$^2$. It is seen that, neglecting other sources of errors like, for instance, the calibration 
of the  lattices, but being conservative with the choice of the fitting window, one can 
conclude that our best-fit parameters incorporating only~\footnote{we define the errors by 
taking the larger ellipsoid and this could be maybe considered as to give account of some systematic effect 
related to the choice of the fitting window.} statistical errors are:
\beq\label{best-fit}
\Lambda_{\overline{\rm MS}}^{Nf=0 }&=&224^{+8}_{-5} \ \rm{MeV} \nonumber \\
g^2_T \langle A^2 \rangle_R &=& 5.1^{+0.7}_{-1.1} \ \rm{GeV}^2 \ .
\eeq
These values are in very good agreement with the previous estimates
from quenched lattice simulations of the three-gluon Green 
function~\cite{OPEtree,OPEone} or, in the case of $\Lambda_{\overline{\rm MS}}$, 
from the implementation of the Schr\"odinger functional method~\cite{Luscher:1993gh}, although 
slightly larger than the one obtained by the ratio of ghost and gluon 
dressing functions~\cite{Boucaud:2005xn} (see fig.~\ref{chi2eps}.(b) and tab.~\ref{comp}). 
Concerning the gluon condensate estimate only, it is worth pointing that it can be computed 
at the renormalization momentum $\mu^2=100$ GeV$^2$ (see tab.~\ref{comp}) and it also agrees very 
well with the estimate from the analysis of the quark propagator vector part, $Z_\psi$, that 
gives: $\sqrt{\langle A^2 \rangle_{R,\mu=10 \ \rm{GeV}}}= 1.76(8)$ GeV~\cite{Boucaud:2005rm}.

\begin{table}
\begin{center}
\begin{tabular}{|c||c|c|c|c|c|}
\hline
& $F^2G$ (this work) & Asym. 3-g \cite{OPEone} & Sym. 3-g \cite{OPEone} &  $F/G$~\cite{Boucaud:2005xn} & \cite{Luscher:1993gh} \\ \hline
$\Lambda_{\overline{\rm MS}}$ (MeV) & 224$^{+8}_{-5}$ &  260(18) & 233(28) & 270(30) & 238(19) \\
\hline
$\sqrt{\langle A^2 \rangle_{R,\mu}}$ (GeV) & 1.64(17) & 2.3(6) & 1.9(3) & 1.3(4) & -- \\
\hline
\end{tabular}
\end{center}
\caption{\small Comparison of estimates of $\Lambda_{\overline{\rm MS}}$ obtained from the analysis of 
the ghost-gluon vertex in this work (first column), the asymmetric 3-gluon vertex (second), the symmetric 3-gluon vertex 
(third), the ratio of gluon and ghost dressing functions (fourth) and 
with the Schr\"odinger functional method (last). The gluon condensate $\langle A^2 \rangle_{R,\mu}$ has 
been obtained at the renormalization momentum $\mu=10$ GeV, for the sake of comparison with the other 
estimates, from \eq{best-fit} by applying $g^2(\mu^2=100 \ \rm{GeV}^2)/4 \pi=0.15$.}
\label{comp}
\end{table}

As a final remark, had we taken into account the leading-logarithm behaviour 
of the Wilson coefficient for the running coupling (applied \eq{ap-alphahNP}) 
instead of \eq{alphahNP}~), the parameters so fitted would not significantly 
differ from those in \eq{best-fit}: we estimate a difference of $\sim 4 \%$ in 
the determination of $g^2_T \langle A^2 \rangle_R$ and 
less than $0.5 \%$ in that of $\Lambda_{\overline{\rm MS}}$.

\section{Conclusions}

In the present paper we reconsider in some detail the determination
of $\Lambda_{\overline {\rm MS}}$ from gluon and ghost Green functions using 
the MOM scheme. We stick here to the quenched case, or rather to the pure
Yang-Mills $SU(3)$ theory,  having of course in mind to
apply what we learn also to the unquenched situation.

\subsection{ghost-gluon vertex}
We give some details about the proper renormalisation of the ghost-gluon vertex
in the MOM scheme mainly because we realised that there is some carelessness in
literature. An obvious remark is that applying  MOM to a vertex function needs
to specify the kinematics of the renormalisation point. Renormalising at the
scale $\mu$ may be performed in the symmetric case, with the three momenta at
the renormalisation scale ($p^2=\mu^2$) or in the soft gluon limite
($p_{gluon}=0, p^2_{ghost}=\mu^2$), or with a vanishing incoming ghost 
momentum, etc. The latter case is the one in which Taylor's theorem applies
which leads to $\widetilde Z_1=1$. We present in section \ref{PTh} an
alternative derivation of the perturbative renormalisation of the coupling
constant in the latter scheme, defined by eq.~ \eq{alpha}, in agreement with the
result  by Chetyrkin~\cite{Chetyrkin00}. The other kinematics  lead to a finite
but non trivial $\widetilde Z_1=1+O(\alpha^2)$. This difference has been often 
overlooked, presumably because it is assumed to be small. However, as we have
shown  in section \ref{PTh}, applying $\widetilde Z_1=1$ to the symmetric case leads to a 14
\%  systematic error on $\Lambda_{\overline {\rm MS}}$ while it gives 23 \% when
applied to the soft gluon limit.

\subsection{The $\Lambda_{\overline {MS}}$ plateau}
 
 $\Lambda_{\overline {\rm MS}}$ is a constant independent on the scale $\mu$. 
Inverting the perturbative expansion of the coupling constant one can 
invert eq.~\eq{betainvert} leading for each $\mu$ to 
$\Lambda_{\overline {\rm MS}}(\mu^2)$ from $\alpha_T(\mu^2)$~\footnote{This can
be done in any MOM scheme using the apropriate equivalent to
\eq{betainvert}. }. If we were in a perturbative region of $\mu$ 
$\Lambda_{\overline {\rm MS}}(\mu^2)$ should not depend on $\mu$ up to
statistical errors.  One should see a nice "plateau''. 
Fig.~\ref{plot-plateau}.(a) shows that this is far from being the case up to
$\mu^2=30 \,{\mathrm GeV}^2$. We have since long advocated that there is a
sizeable non-perturbative contribution from the vev of the unique (in Landau
gauge) dimension 2 operator $\langle A^2 \rangle$. We propose to fit this
condensate by adjusting  the resulting $\Lambda_{\overline {\rm MS}}$ to a
``{\it plateau}''. This is  successfully achieved, see
fig.~\ref{plot-plateau}.(b). Since we scan a large window in the scale $\mu$ 
we believe that we are in a position to claim that we indeed see a non-perturbative
$O(1/\mu^2)$ contribution rather than the effect of logarithmically behaved higher orders in
perturbation theory ($O(\alpha^5)$). 

\subsection{Comparison of different estimates of $\Lambda_{\overline {MS}}$}
We have performed a comparison of different estimates of $\Lambda_{\overline
{\rm MS}}$ and $\langle A^2 \rangle$ in the pure Yang-Mills theory using the
coupling constant defined in \eq{alpha}, the MOM coupling constant from
symmetric three gluon vertex function, the MOM coupling constant from the three
gluon vertex function with one vanishing momentum and from the ghost to gluon
propagator ratio, and also with the estimate of $\Lambda_{\overline {\rm MS}}$
from the Schr\"odinger functional approach. The result is reported in table
\ref{comp} and fig. \ref{chi2eps}.(b). The agreement is quite satisfactory. 
Fig~\ref{plot-alpha}.(b) shows also a good agreement of our fit from  
$\alpha_T(\mu^2)$ with very large $\mu$ measurements
from~\cite{Sternbeck:2007br}. Notice also that $\Lambda_{\overline {\rm MS}}$
from $\alpha_T(\mu^2)$ has the smallest statistical errors due to the fact that
it relies only on propagator, not on noisier three point Green functions.  

This opens a possibility of using the matching of $\Lambda_{\overline{\rm MS}}$ 
as computed from different lattices in order to fit the lattice spacing ratio.
One might also match directly $\alpha_T(\mu^2)$ from different lattices, 
a procedure which is not constrained to large scales and does not need to 
estimate the $ \langle A^2 \rangle$ condensate. In fact from \eq{alpha} we get 
directly a quantity which should be independant of the lattice spacing at
the same $\mu$ in physical units, up to 
$O(1/a^2)$ artifacts. This method is complementary to the use of Sommer's
parameter $r_0$~\cite{Sommer:1993ce} and it also only depends on gauge fields.

\appendix

\section{Appendix: ghost and gluon propagators anomalous dimension in MOM}
\label{appendix}

The ghost and gluon anomalous dimension can be computed 
in MOM scheme by applying \eq{eq:evolution} with the results 
obtained in ${\overline{\rm MS}}$ for the radiative corrections 
of all the relevant Green functions~\cite{Chetyrkin00,Chetyrkin:2004mf}. 
Thus, one obtains for the coefficients 
defined in \eq{gammas} :

\beq
\widetilde{\gamma}_0 &=& \frac {9} 4 \nonumber \\
\widetilde{\gamma}_1 &=& \frac{813}{16}-\frac{13 N_f}{4} \nonumber \\
\widetilde{\gamma}_2 &=& \frac{157303}{64}-\frac{14909 N_f}{48}+
\frac{125 N_f^2}{18}-\frac{5697 \zeta(3)}{32}
-\frac{21}{4} N_f \zeta(3) \nonumber \\
%
\widetilde{\gamma}_3 &=& \frac{219384137}{1536}-\frac{30925009 N_f}{1152}
+\frac{288155 N_f^2}{216}-\frac{2705 N_f^3}{162}
-\frac{9207729 \zeta(3)}{512} \nonumber 
\\ && + \frac{132749}{96}
N_f \zeta(3)-\frac{19}{2} N_f^2 \zeta(3)
-\frac{221535 \zeta(5)}{32}+\frac{15175}{16} N_f \zeta(5)
\eeq

\beq
\gamma_0 &=& \frac{13}{2}-\frac{2 N_f}{3}
\nonumber \\ 
\gamma_1 &=& \frac{3727}{24}-\frac{250 \ N_f}{9}+\frac{20 \ N_f^2}{27}
\nonumber \\
\gamma_2 &=& \frac{2127823}{288}-\frac{9747 \ \zeta(3)}{16}
+ N_f \left(-\frac{5210}{3}+\frac{119 \zeta(3)}{3}\right) 
\nonumber \\
&&+N_f^2 \left(\frac{1681}{18}+\frac{16 \ \zeta(3)}{9}\right)
-\frac{200 \ N_f^3}{243} 
\nonumber \\
\gamma_3 &=& \frac{3011547563}{6912}-\frac{18987543
\zeta(3)}{256}-\frac{1431945 \zeta(5)}{64}
\nonumber \\
&& +N_f
\left(-\frac{221198219}{1728}+\frac{2897113 \zeta(3)}{216}+\frac{845275 \zeta(5)}{96}\right)
\nonumber \\
&& +N_f^2 \left(\frac{6816713}{648}-\frac{60427 \zeta(3)}{162}-\frac{4640 \zeta(5)}{9}\right) 
\nonumber \\
&& + N_f^3 \left(-\frac{373823}{1458}-\frac{88 \zeta(3)}{27}\right)
+\frac{2000 N_f^4}{2187}
\eeq
These coefficients appear for the expansion, given by \eq{eq:evolution}, 
of the MOM-renormalized ghost and gluon anomalous dimension in terms of the 
$\overline{\rm MS}$-coupling. However, provided that the $\beta$-function 
for any other renormalization scheme is known, it can be applied 
to replace $\alpha_{\overline{\rm MS}}$ in \eq{gammas} by the coupling 
in that scheme.

\section{Appendix: Wilson coefficients at leading logarithms}
\label{appendix2}

The purpose of this appendix is to present up to leading logarithms the subleading 
Wilson coefficients in eqs.~(\ref{Z3fantome},\ref{Z3glue}) and, in view of  
checking the validity of neglecting those logarithms, estimate their impact on 
the momenta window we use for our fits.
Following \cite{OPEone}, let us write
\beq\label{ap-props}
G_R(q^2,\mu^2) &=& c_0\left(\frac{q^2}{\mu^2},\alpha(\mu^2)\right) + 
c_2\left(\frac{q^2}{\mu^2},\alpha(\mu^2)\right) \ \frac{\langle A^2_R \rangle_\mu}{4 (N_c^2-1) q^2}
\nonumber  \\
F_R(q^2,\mu^2) &=& \widetilde{c}_0\left(\frac{q^2}{\mu^2},\alpha(\mu^2)\right) + 
\widetilde{c}_2\left(\frac{q^2}{\mu^2},\alpha(\mu^2)\right) \ \frac{\langle A^2_R \rangle_\mu}{4 (N_c^2-1) q^2} \ 
\eeq
for gluon and ghost propagators. Then, with the help of the appropriate renormalization constants
one can rewrite \eq{ap-props} in terms of bare quantities:
\beq\label{ap-bar}
G(q^2,\Lambda^2) &=& Z_3(\mu^2,\Lambda^2) \ c_0\left(\frac{q^2}{\mu^2},\alpha(\mu^2)\right)
\nonumber \\ 
&+&  Z_3(\mu^2,\Lambda^2)  Z_{A^2}^{-1}(\mu^2,\Lambda^2) \ c_2\left(\frac{q^2}{\mu^2},\alpha(\mu^2)\right) \
\frac{\langle A^2 \rangle}{4 (N_c^2-1) q^2} \ ,
\eeq
where $A^2_R=Z^{-1}_{A^2} A^2$.
A totally analogous equation for the ghost dressing function $F(q^2,\Lambda^2)$, 
with $\widetilde{c}_i$ and $\widetilde{Z}_3$ in place of $c_i$ and $Z_3$. 
Now, as the $\mu$-dependence of both l.h.s. and r.h.s. of \eq{ap-bar} should match 
each other for any $q$, one can take the logarithmic derivative with respect to $\mu$ and 
infinite cut-off limit, term by term, on r.h.s. and obtains: 
\beq\label{ap-diffeqs}
\gamma(\alpha(\mu^2)) + 
\left\{ \frac{\partial}{\partial\log\mu^2} 
+ \beta(\alpha(\mu^2))\frac{\partial}{\partial \alpha}\right\} \ 
\ln c_0\left(\frac{q^2}{\mu^2},\alpha(\mu^2)\right) &=& 0 
\nonumber \\
-\gamma_{A^2}(\alpha(\mu^2)) + \gamma(\alpha(\mu^2)) 
+ \left\{ \frac{\partial}{\partial\log\mu^2} 
+ \beta(\alpha(\mu^2))\frac{\partial}{\partial \alpha}\right\} \ 
\ln c_2\left(\frac{q^2}{\mu^2},\alpha(\mu^2)\right) &=& 0 \ ,
\eeq
where $\gamma(\alpha(\mu^2))$ is the gluon propagator anomalous 
dimension defined in \eq{gammas} and 
\beq
\gamma_{A^2}(\alpha(\mu^2)) \ = \ 
\lim_{\Lambda \to \infty} \ \frac{d}{d\ln\mu^2} \ln Z_{A^2}(\mu^2,\Lambda^2) \ 
= \ - \gamma_0^{A^2} \ \frac{\alpha(\mu^2)}{4 \pi} \ + \dots 
\eeq
Both eqs.~(\ref{ap-diffeqs}) can be finally combined to give:
\beq\label{ap-fin}
\left\{ -\gamma_{A^2}(\alpha(\mu^2)) +  \frac{\partial}{\partial\log\mu^2} 
+ \beta(\alpha(\mu^2))\frac{\partial}{\partial \alpha}\right\} \ 
\frac{c_2\left(\frac{q^2}{\mu^2},\alpha(\mu^2)\right)}
{c_0\left(\frac{q^2}{\mu^2},\alpha(\mu^2)\right)} \ = \ 0 \ .
\eeq

\begin{figure}[htb]
\begin{center}
\includegraphics[width=8.5cm]{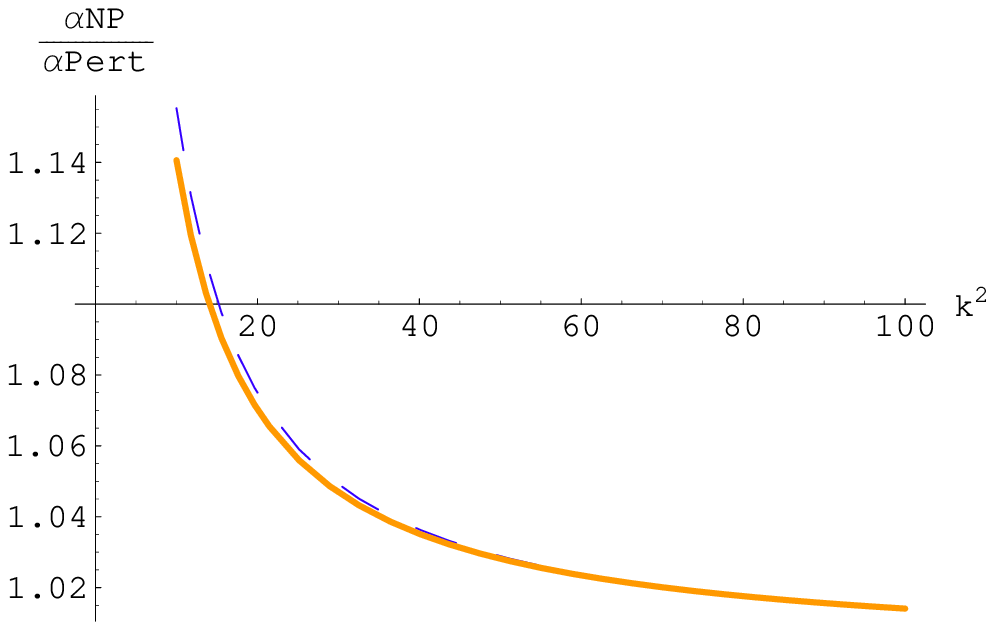}
\end{center}
\caption{\small $\alpha^{\rm NP}/\alpha^{\rm pert}$ in terms of the square of the momentum 
computed by using both \eq{ap-alphahNP} (dashed blue) and \eq{alphahNP} (solid red).}
\label{ap-fig}
\end{figure}

We can proceed in the same way for the ghost dressing function and derive analogous 
equations for the Wilson coefficients, $\widetilde{c}_i$, that differ from those 
for $c_i$ only because $\widetilde{\gamma}(\alpha(\mu^2))$ 
takes the place of $\gamma(\alpha(\mu^2))$. Thus, the combination 
$\widetilde{c}_2/\widetilde{c}_0$ obeys exactly the same \eq{ap-fin}, above derived 
for $c_2/c_0$, that can be solved at the leading logarithm
as explained in \cite{OPEone} to give:
\beq\label{ap-sols}
\frac{c_2\left(\frac{q^2}{\mu^2},\alpha(\mu^2)\right)}
{c_0\left(\frac{q^2}{\mu^2},\alpha(\mu^2)\right)}
\ = \
\frac{\widetilde{c}_2\left(\frac{q^2}{\mu^2},\alpha(\mu^2)\right)}
{\widetilde{c}_0\left(\frac{q^2}{\mu^2},\alpha(\mu^2)\right)}
\ = \ 3 g^2(q^2) \ \left( \frac{g^2(q^2)}{g^2(\mu^2)}\right)^{-\gamma_0^{A^2}/\beta_0}
\ .
\eeq
The boundary condition comes from requiring \eq{ap-diffeqs} to be equal to 
\eq{Z3fantome} for the ghost and \eq{Z3glue} for the gluon at $\mu^2=q^2$.
The coefficient $\gamma_0^{A^2}$ was computed to be 35/4 for the first time in \cite{OPEone}.
Of course, eqs.~(\ref{ap-diffeqs}) define not only the dependence of the Wilson coefficient 
on the renormalization momentum, $\mu^2$, but also that on the momentum scale $q^2$ because 
of standard dimensional arguments: the only dimensionless quantities\footnote{Other 
dimensionless quantities can be obtained with the help of $\Lambda_{QCD}$, but this 
is a non-perturbative parameter not emerging in the Wilson coefficient dominated by 
the short-distance singularities of the OPE expansion and only coding perturbative information
in the SVZ approach.} are the ratio $q^2/\mu^2$ and $\alpha$. 
Then, putting all toghether, the non-perturbative formula for the running coupling 
at the leading logarithm is given by 
\beq\label{ap-alphahNP}
\alpha_T(\mu^2)
\ = \
\alpha^{\rm pert}_T(\mu^2)
\ 
\left( 
 1 + \frac{9}{\mu^2} 
\left( 
\frac{\ln\frac{\mu^2}{\Lambda^2_{QCD}}}{\ln\frac{\mu_0^2}{\Lambda^2_{QCD}}}
\right)^{-9/44}
\frac{g^2_T(\mu_0^2) \langle A^2 \rangle_{R,\mu_0^2}} {4 (N_C^2-1)}
\right) \ ,
\eeq
where the only correction to \eq{alphahNP} comes from the 
ratio of logarithms inside the bracket that, as can be seen in 
fig.~\ref{ap-fig}, introduces no significant deviation. 

\end{document}